\documentstyle[prl,twocolumn,aps,epsf]{revtex}

\begin{document}
 \tolerance 50000
\def\d{\dagger}
\draft

\twocolumn[\hsize\textwidth\columnwidth\hsize\csname@twocolumnfalse\endcsname

\title{ Alternating-Spin  Ladders in a Magnetic Field: New Magnetization Plateaux}

\author{A.~Langari$^{\dagger}$  and M.A. Mart\'{\i}n-Delgado$^{\ast}$  
} 
\address{
$^{\dagger}$ Max-Planck-Institut f\"ur Physik komplexer Systeme, N\"othnitzer Strasse 38,
D-01187 Dresden, Germany
\\
$^{\ast}$Departamento de F\'{\i}sica Te\'orica I, Universidad Complutense,
28040-Madrid, Spain.  
}

\maketitle 

\begin{abstract}
We study numerically the formation of magnetization plateaux with the 
Lanczos method in 2-leg ladders with mixed spins of magnitudes
$(S_1,S_2)=(1,1/2)$ located at alternating positions along the 
ladder and with dimerization $\gamma$.
For interchain coupling $J'>0$ and $\gamma=0$,
we find normalized plateaux at $m=1/3$
starting at zero field and $m=1$ (saturation), while when 
$\gamma \ne 0$ is columnar, another extra plateau at $m=2/3$ shows up. 
For $J'<0$, when
$\gamma<\gamma_c(J')$ we find no plateau  while for $\gamma>\gamma_c(J')$
we find four plateaux at $m=0,1/3,2/3,1$. We also apply several approximate
analytical methods (Spin Wave Theory, Low-Energy Effective Hamiltonians
 and Bosonization)
to understand  these findings and to conjeture the behaviour 
of ferrimagnetic ladders with a bigger number of legs.
\begin{center}
\parbox{14cm}{}

\end{center}
\end{abstract}

\vspace{-0.8 true cm}

\pacs{
\hspace{2.5cm} 
}

\vskip2pc]
\narrowtext


\vspace {0.5 cm}
\section{Introduction}

Low dimensional spin systems has attracted many interests in recent years.
It is the dominant effect of quantum fluctuation which 
distinguishes them from the
classical picture which is valid for higher dimensions. 
For instance it was argued by
Haldane\cite{Haldane} that integer spin chains have an energy gap between
the ground state and the first excited state. 
In the presence of magnetic field, the energy gap persists 
up to a critical field, equal to the gap. For $S$=1 
Anti-Ferromagnetic Heisenberg (AFH) chain 
the spectrum is gapless from the critical field up to the saturation 
field\cite{Sakai}. 
The $S$=1/2 AFH chain remains gapless from
zero field up to the saturation field, where the ground state is fully
polarized\cite{Griffiths}.  The situation might be more complex 
when one considers  non-nearest neighbour interactions
or non-homogeneous spin system by considering Bond-Alternations (BA) 
or different kind of spins. However a similar information may be obtained
about the full spectrum by study the magnetization of system in the
presence of a magnetic field.

The study of the magnetization process in one-dimensional quantum spin systems
has been spured since the discovery of quantized  plateaux in the magnetization
curves as a function of the applied magnetic field 
\cite{hida},\cite{OYA},\cite{totsuka}. 
This behaviour is a sort of ``magnetic analogy'' of the Quantum Hall Effect
for charge degrees of freedom and  it has become an issue of both 
theoretical and experimental interest.
Likewise, 1d-chains with alternating mixed spins have been shown to behave
with unusual quantum properties: they show ferrimagnetic behaviour
instead of antiferromagnetism \cite{Kriv}-\cite{rd}.
For instance for low spin values of ($S_1, S_2$) ferrimagnetic chain 
($S_1 \neq S_2$)
the low energy spectrum is gapless while  there is a gap above this spectrum.
Thus, a natural question arises as to whether ferrimagnetic systems
are on equal footing with AF systems as far as the formation of 
magnetization plateaux is concerned.
Here a remark is in order. As an alternating spin chain has an ordered
ground state (GS) with total spin $S={N\over2}(S_1-S_2)$, 
then a natural plateau is
expected even at zero field. Being this GS magnetization quite robust,
a possible scenario for quantum ferrimagnets is that no other plateaux
will appear, in addition to the saturation one. We shall show that this
is not the case.

\noindent Although there are recent studies of one-chain ferrimagnets
with magnetic fields \cite{maisinger},\cite{sakai-yamamoto}, 
to our knowledge little is known about 
the onset of magnetization plateaux in ferrimagnetic ladders.
A ferrimagnetic ladder may be regarded as the first step to consider
the interaction between ferrimagnetic chains as shown in Fig.1.
This structure  exists in $MnCu(pbaOH)(H_2O)_3$
where it is composed of parallel planes containing 
ferrimagnetic chains and in each  
plane the interaction between chains is assumed to be smaller 
than the intra-chain interactions\cite{Kahn}.
In this paper we address the study of ferrimagnetic ladders under 
several situations of the 
Anti- ($J'>0$) and Ferro-magnetic ($J'<0$) interchain coupling 
and $\gamma$ dimerization patterns,
which reveal a very reach structure of plateaux at fractional values
of the full magnetization.

\begin{figure}
\epsfxsize=8.5cm \epsfysize=7cm  \epsffile{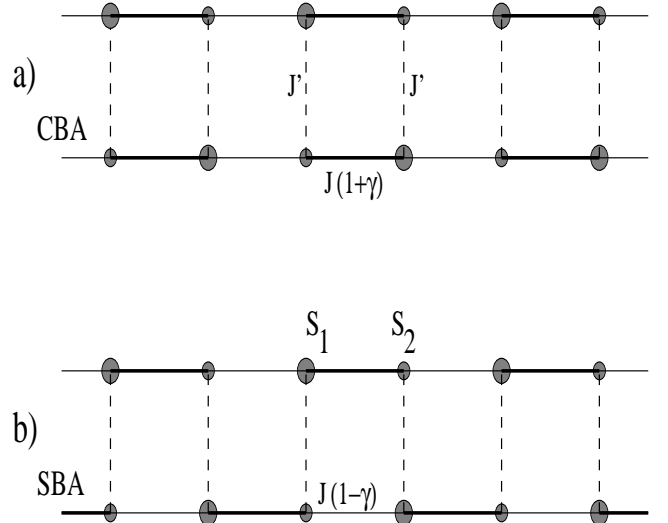}
\narrowtext
\caption[]{ A picture of the dimerization patterns 
considered in a 2-leg mixed spin ladder: a) Columnar Bond Alternation (CBA),
b) Staggered Bond Alternation (SBA).
}
\label{fig0} 
\end{figure}

We define the total magnetization $M$ per unit length  as follows,

\begin{equation}
M = {1\over L} \sum_{\alpha=1}^2 \sum_{n=1}^L \langle S_1^{(\alpha)z}(n) + 
S_2^{(\alpha)z}(n) \rangle \;,
\label{0}
\end{equation}

\noindent where ${\bf S}^{(\alpha)}_1(n)$ denotes the quantum spin-$1$ 
at site $n$ in the leg $\alpha=1,2$ of the ladder, similarly 
${\bf S}^{(\alpha)}_2(n)$ for the spin-$1/2$   and 
the number of sites is $N=2\times L$,  $L$ is the number of rungs.
Alternatively we may use $m$, the normalized value of the magnetization
with respect to the saturation value as $m=M/M_{sat}$.
We want to study this observable in a 2-leg ferrimagnetic ladder with the 
following Heisenberg Hamiltonian ($i=2n+\alpha$),

\begin{eqnarray}
H &=& J \sum_{\alpha=1}^2 \sum_{n=1}^{{L\over 2}-1} \; (\;
[1+\gamma^{(\alpha)}(i)]
\; \; {\bf S}^{(\alpha)}_1(i) \cdot  {\bf S}^{(\alpha)}_2(i+1) \nonumber \\ 
&+&[1 + \gamma^{(\alpha)}(i+1)]\;\;
{\bf S}^{(\alpha)}_2(i+1) \cdot  {\bf S}^{(\alpha)}_1(i+2)\; ) \nonumber \\
& + & J^\prime \sum_{\alpha \neq \beta}^{1,2}\sum_{n=1}^L
{\bf S}^{(\alpha)}_1(n) \cdot {\bf S}^{(\beta)}_2(n) 
- h S^z_{total} \;,
\label{1}
\end{eqnarray}

\noindent where $h$ is the magnetic field,  
$J$ and $J^\prime$ are coupling constants 
along the legs and the rungs respectively, and $\gamma^{(\alpha)}(n)$
is the dimerization patterns. We consider two different dimerization
patterns : 
i) CBA (Columnar Bond Alternation, Fig.\ref{fig0}-(a))
for which $\gamma^{(\alpha)}(n)=(-1)^{n+1}\gamma$ \cite{QRG}, 
and ii) SBA (Staggered Bond Alternation, Fig.\ref{fig0}-(b)) where
$\gamma^{(\alpha)}(n)=(-1)^{\alpha+1+n}\gamma$ \cite{QRG}.
We use periodic boundary conditions along the legs of the ladder.

The paper is organized as follows. In the next section we show the application
of spin wave theory to derive the saturation field. In Sec. III we obtain
a low-energy effective Hamiltonian based on Quantum Renormalization Group (QRG)
formalism to describe the magnetization curve and give a conjecture for 
a $2n$-leg ferrimagnetic ladder. We present an Abelian Bosonization 
in Sec. IV to support our conjecture and express the necessary condition
for the appearance of the magnetization plateaux in a $2n$-leg ferrimagnetic
ladder. In Sec. V we will confirm the results of the analytical methods
using a numerical method (Lanczos method) applied to a 2-leg ladder under a
variety of situations. 
Finally, we summerize our results as a conclusion in Sec. VI.


\section{ Spin Wave Theory} We start considering the case $J'>0$.
Then, since the GS is an ordered state with non-vanishing 
magnetization, it is natural to  apply SWT to study the low energy 
properties of these ladders  and predict correctly their ferrimagnetic
behaviour. 
Now we extend this analysis including the
effect of the external magnetic field.
For weak magnetic fields we expect that the ferrimagnetic GS will be 
little perturbed so that it is still a good choice to make a SWT analysis
around a partially polarized state with magnetization 
$S^z_{GS}={N\over 2} (S_1-S_2)$.
At $h=0$ the 2-leg ladder presents two pairs of branches, say
$\omega_A^{(\pm)}(k;h=0), \omega_B^{(\pm)}(k;h=0)$, 
in the dispersion relation, namely:
\begin{eqnarray}
\omega_A^{(\pm)}(k;0)=\frac{1}{2}(C_1-C_2+\sqrt{(C_1+C_2)^2-4C_3^{2 (\pm)}(k)}),
\nonumber \\
\omega_B^{(\pm)}(k;0)=\frac{1}{2}(C_2-C_1+\sqrt{(C_1+C_2)^2-4C_3^{2 (\pm)}(k)}),
\label{2}
\end{eqnarray}
\begin{eqnarray}
C_1=(2J+J')S_1 \; \; \;; \; \; \; C_2=(2J+J')S_2  \;,\\
C_3^{(\pm)}(k)=\sqrt{S_1S_2}(2J \cos k \pm J') \;.
\label{3}
\end{eqnarray}

\noindent
For  $J'=0$ the two chains decouple and the branches collapse by pairs
reproducing the one-chain result. When $h$ is applied the spectrum (\ref{2}) is
modified as follows:

\begin{eqnarray}
\omega_A^{(\pm)}(k;h) = \omega_A^{(\pm)}(k;h=0) - h \;,
\nonumber \\
\omega_B^{(\pm)}(k;h) = \omega_B^{(\pm)}(k;h=0) + h \;,
\label{4}
\end{eqnarray}
\noindent After doing this analysis we find that the Bogoliubov
transformation to diagonalize the SWT Hamiltonian is 
{\em independent of $h$}. This fact implies that we cannot see
the plateaux of the curves $M=M(h)$ using this SWT.

For strong magnetic fields the previous ferrimagnetic GS is not a good choice
to perturb around. It is more natural to choose a fully polarized
(ferromagnetic) state with total magnetization $S^z_{GS}={N\over 2} (S_1+S_2)$.
We find that the SWT analysis yields the following spectrum:

\begin{eqnarray}
\Omega_A^{(\pm)}(k;0)&=&\frac{1}{2}(-(C_1+C_2) \pm 
\sqrt{(C_1+C_2)^2-D^{(+)}(k)}),
\nonumber \\
\Omega_B^{(\pm)}(k;0)&=&\frac{1}{2}(-(C_1+C_2) \pm  \sqrt{(C_1+C_2)^2-D^{(-)}(k)}), \nonumber \\
D^{(\pm)}(k)&=&16 J S_1S_2 (J \pm J \cos k + J')(1 \mp \cos k).
\label{5}
\end{eqnarray}

\noindent The inclusion of the magnetic field merely shifts this spectrum
by $h$:

\begin{eqnarray}
\Omega_A^{(\pm)}(k;h) = \Omega_A^{(\pm)}(k;h=0) + h\;,
\nonumber \\
\Omega_B^{(\pm)}(k;h) = \Omega_B^{(\pm)}(k;h=0) + h\;.
\label{6}
\end{eqnarray}
\noindent Again, we are not able to detect magnetization plateaux.
However, now we can give a reliable prediction for the location of
the saturation field $h_{sat}$. This happens when 
$\Omega_A^{(-)}(k=0;h_{sat}) \equiv 0$, giving us the following 
relation for $S_1=1, S_2=1/2$ and normalizing $J=1$:

\begin{equation} 
h^{sat} = {3\over2} (2+J').
\label{7}
\end{equation}

\noindent We shall find that this condition is verified numerically.
For decoupled chains $J'=0$ we reproduce the one-chain results 
\cite{sakai-yamamoto}.
Similarly, for ferromagnetic coupling  $J'<0$ we find
the following saturation field,

\begin{equation}
h_{sat} = {3\over4}
\left( (2+J') + \sqrt{(2+J')^2-{64J'\over9}}\right). 
\label{14}
\end{equation}


\begin{figure}
\epsfxsize=8.5 cm \epsfysize=7cm  \epsffile{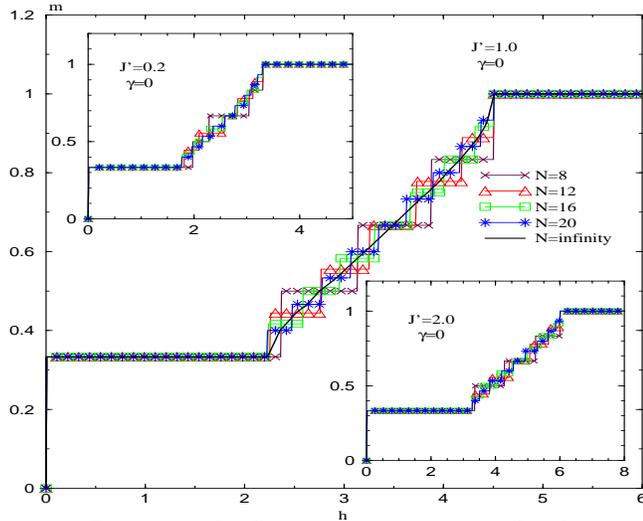}
\narrowtext
\caption[]{ Magnetization plateaux for alternating 2-leg ladder:
$\gamma=0$, $J'=0.2, 1.0, 2.0$. There are two plateaux at $m=1/3\;,\;1$.
}
\label{fig1} 
\end{figure}
\noindent

\section{Low Energy Effective Hamiltonian} 

In order to explain qualitatively
the onset of magnetization plateaux in a 2-leg ferrimagnetic ladder
we can use an RG transformation to map the original model (\ref{1})
onto a one-chain model with known properties. 
To perform the block RG transformation we choose as the building block
the unit cell made up of 2 spins $(S_1=1,S_2=1/2)$. The block Hamiltonian
is then $h^{B}=J' {\bf S}_1\cdot{\bf S}_2 - h (S_1^z+S_2^2)$.
For simplicity, we assume a strong coupling regime $J'\gg J$.
This allows us to place $h^B$ on the vertical 
rungs of the 2-leg ladder so that 
the whole block Hamiltonian is  the rung Hamiltonian ($H^B=H_r$) and the 
inter-block Hamiltonian is   the leg Hamiltonian ($H^{BB}=H_l$).
We need to distinguish 2 regimes of magnetic field to construct the 
truncation operator. 
\noindent
a) For low fields, namely $h<3/2 J'$, the lowest lying 
states of $h^B$ is a spin-1/2 doublet. Then we use this to map (\ref{1})
onto the following low-energy effective Hamiltonian ($J=1$):

\begin{eqnarray}
H_{eff} = -LJ' -&{4\over9}&\sum_{n=1}^L \gamma'_n \ {\bf S}'_n\cdot{\bf S}'_{n+1}
-h\sum_{n=1}^L S'^z_n \;,\nonumber \\
%
\gamma'_n &=& \sum_{\alpha=1}^2 (1+\gamma^{(\alpha)}(n))\;,
\label{9}
\end{eqnarray}

\begin{figure}
\hspace{-0.8cm}
\epsfxsize=8.5cm  \epsfysize=7cm \epsffile{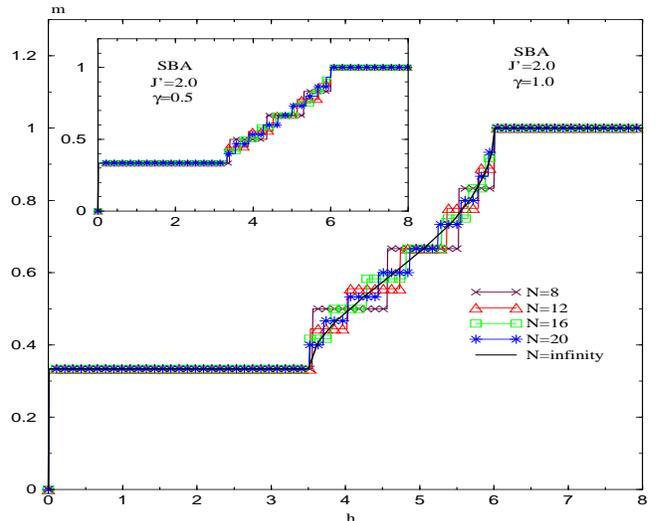}
\narrowtext
\caption[]{Magnetization curves for SBA configuration with $J'=2.0$,
$\gamma=1.0, 0.5$ which shows only two plateaux at $m=1/3\;,\;1$.
}
\label{fig2} 
\end{figure}
\noindent

\noindent $H_{eff}$ is a spin (${\bf S}'=$) 1/2 Ferromagnetic Heisenberg chain in the
presence of an external magnetic field, which shows a ferromagnetic phase
with $M=1/2 \ (m=1/3)$.

\noindent 
b) For high fields, namely $h>3/2 J$, the lowest two energy states of $h^B$
have now spin quantum numbers $|S_T=1/2,S^z_T=+1/2\rangle $, $|3/2,+3/2\rangle$.
Upon projection of (\ref{1}) onto this subspace we arrive at the following
effective Hamiltonian ($J=1$):

\begin{equation}
H_{eff}= -({J'\over 4}+h)L + 
{5\over 36} \sum_{n=1}^L \gamma'_n + H_{XXZ+h} \;,
\label{10}
\end{equation}

\noindent where $H_{XXZ+h}$ is an AF spin 1/2 Heisenberg model in a magnetic field
with coupling parameters given by:

\begin{eqnarray}
H_{XXZ+h}&=& 
{2\over3}\sum_{n=1}^L  
\gamma'_n(\sum_{\mu=x,y}S_n^{\prime\mu}S^{\prime\mu}_{n+1} +
\Delta S_n^{\prime z}S^{\prime z}_{n+1})
-h'_n S^{\prime z}_n  \;,\nonumber \\
\Delta &=& {1\over3}\; \; \; \; ; \; \; \; \;  h'_n = {3\over2}h - 
{11\over6}\gamma'_n -{9\over4}J' \;,
\label{12}
\end{eqnarray}

\noindent In the absence of dimerization $\gamma=0$ or 
when the dimerization pattern is SBA,
we have $\gamma'_n=2$  constant and thus the effective field $h'$ is also uniform.
In this case we can use the knowledge of the $H_{XXZ+h}$ model to predict the existence of 2 plateaux.  It is known that this model has a critical value for $h'$ where a phase
transition occurs between the Partially Magnetized (PM) to the Saturated Ferromagnetic
(SF) states. This critical line in the $\Delta-h'$ phase diagram is given by
\cite{yang}  $h'_c=\Delta+1$ (for $h'>0$).
Firstly, we find that the relation between the magnetization $M$ of the
original ferrimagnetic ladder and the magnetization $M_{XXZ}$ of the effective XXZ model
is:

\begin{equation}
M = 1 + M_{XXZ}.
\label{13a}
\end{equation}

\noindent Now it is possible to derive the dependence of $M$ on $h$. 
By using the relation in (\ref{12}) we find
two critical fields given by $h_{c1}={2\over3}+{3\over2}J'$ and $h_{c2}={38\over9}+{3\over2}J'$ such that for $h\le h_{c1}$ then $M=1/2$, for 
$h_{c1}\le h \le h_{c2}$ then $1/2\le M \le 3/2$ and for $h\ge h_{c2}$ then
$M=3/2$. Therefore, we find a good qualitative picture of the 
$M(h)$-plot with two plateaux at normalized values of $m=1/3$ and $m=1$
\cite{comment1}.
This agrees with the numerical results in Fig. 1 and 
Fig. 2 using Lanczos.

Although the knowledge of known models obtained in 
Eqs.(\ref{9},\ref{12}) was used to explain the magnetization curve
one may however continue 
using QRG for Eqs.(\ref{9},\ref{12}) to obtain the RG-flow 
of coupling constants and finally the magnetization curve.
In the case of Eq.(\ref{9}) the Hamiltonian will be projected to another
ferromagnetic Heisenberg model with larger spins. By repeating this 
procedure one will arrive at the classical ferromagnetic Heisenberg fixed point.
In the other case in Eq.(\ref{12}), the RG-flow will describe both the
partially magnetized and the fully polarized state of the $XXZ+h$ model
\cite{ab}. 
Needless to say that the final result will not change.

Furthermore, if we restrict ourselves to a strong coupling analysis, we can extend this QRG analysis in a simple way
to  ferrimagnetic ladders with a number of legs given by
$n_l\equiv2n$ in order
to preserve the structure of $(S_1,S_2)$-cells in both directions.
Then, in each rung the highest total spin is $n(S_1+S_2)$ and when we
apply a $h$ field we find that the allowed values of $M$ are 
$-n(S_1+S_2),-n(S_1+S_2)+1,\ldots,n(S_1+S_2)-1,n(S_1+S_2)$. Upon normalization to the highest value, we get that the allowed values of
$m$ are:

\begin{equation}
-1,-1 + {1\over n(S_1+S_2)},\ldots,+1-{1\over n(S_1+S_2)},+1 \;.
\end{equation}

\noindent Now, if we particularize to the case $S_1=1,S_2=1/2$,
we find $-1,-1+2/3n,\ldots,+1-2/3n,+1$. We can readily check that in
this set {\em the values 1/3 and 1 are always present.}
Moreover, we may conjeture the existence of additional plateaux if 
the number of legs increases.


\section{Abelian  Bosonization} 

We may re-obtain this quantization condition using
a heuristic argument based on Abelian Bosonization, as in the original
OYA argument \cite{OYA}. This has been carried out with full detail for
standard AF spin-1/2 ladders in refs. \cite{c-h-p}.
For simplicity, let us consider first a one-chain ferrimagnet ($J'>0$) with
spins $(S_1,S_2)$ in the unit cell and no dimerization $\gamma=0$.
This unit cell will play the role of an effective site in the bosonization.
Firstly, the spin chain is fermionized by introducing colored fermions 
$\psi_n^{(i),c_i}$, $i=1,2$, 
with color index $c_i=1,\ldots,2S_i$ to describe the spin degrees of freedom 
of $S_1$ and $S_2$, respectively. We make the usual assumption about the
continuum limit: only excitations close to the Fermi points are important.
Thus, we introduce slowly varying Left and Right fermionic fields 
$\psi_{L,R}^{(i),c_i}(x)$,  $i=1,2$ such that
$\psi_n^{(i),c_i}\sim e^{ik_Fx}\psi_{R}^{(i),c_i}(x)+
e^{-ik_Fx}\psi_{L}^{(i),c_i}(x)$. Then, the bosonization proceeds in
a standard fashion representing these fermions by vertex operators in
terms of chiral boson fields $\phi_{L,R}^{(i),c_i}$ as 
$\psi_{L,R}^{(i),c_i}(x)=e^{\mp i\phi_{L,R}^{(i),c_i}/R}$, respectively,
and $R$ is the compactification radius of the boson. In this low energy
analysis, all potentially massive modes will decouple and only the 
massless bosons will contribute to this effective description.
This is the case of the total diagonal boson 
$\Phi=\sum_{i=1,2} \sum_{c_i=1}^{2S_i} 
(\phi_{L}^{(i),c_i}+\phi_{R}^{(i),c_i})$. In terms of this boson,
the translational symmetry by one unit cell of the original lattice chain
is realized as a total shift $\Phi \rightarrow \Phi + 2(S_1+S_2)k_F2\pi R$.
On the other hand, the Fermi momentum is modified by a shift of the
magnetization as $k_F={\pi\over2}(1-m)$ (the magnetic field acts as a
chemical potential.)
With these two conditions, a leading operator of the form  $\cos (\Phi/R)$
will be compatible only if $(S_1+S_2)(1-m)=\mbox{integer}$.
This is the application  of the OYA condition to ferrimagnetic systems.

\begin{figure}
\hspace{-0.8cm}
\epsfxsize=8.5cm \epsfysize=8cm  \epsffile{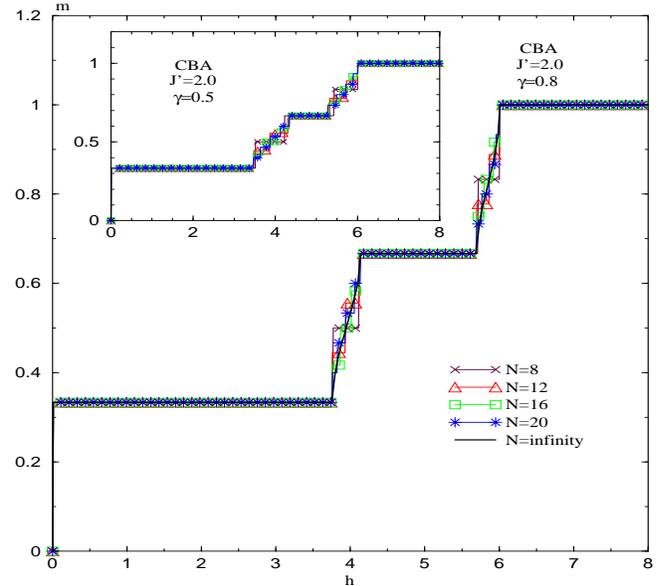}
\narrowtext
\caption[]{Magnetization plateaux  for CBA configuration with $J'=2.0$,
$\gamma=0.8, 0.5$. A new plateau appears at $m=2/3$, its width increases 
by increasing $\gamma$.
}
\label{fig3} 
\end{figure}
\noindent

In our case, we have
$S_1=1,S_2=1/2$ and this yields only $m=1/3,1$.
This agrees with the QRG analysis \cite{comment1} and with numerics
\cite{sakai-yamamoto}.
We may attempt to go beyond this one-chain case to a set of $n$  2-leg 
alternating ladders coupled together. In each 2-leg ladder we assume
a strong coupling limit ($J'\gg 1$) and place the unit cell $(S_1,S_2)$
on the rungs as in the RG analysis. Then, we assume a weak coupling ($J'\ll 1$)
among these set of $n$ ladders. Repeating the argument yields the following
quantization condition for magnetization plateaux:

\begin{equation}
T_u (S_1+S_2) \ n \ (1-m) = \mbox{integer}\;,
\label{13}
\end{equation}

\noindent where we have introduced a ``translational unit'' $T_u$ to
account for the translational symmetry of the ground state.
In the original Hamiltonian we have $T_u=2$ but
due to non-perturbative effects  it can be different
in the real ground state \cite{totsuka}. This condition (\ref{13}) (with $T_u=1$) is precisely equivalent to eq. (\ref{12}) \cite{c-h-p}.
We remark that this condition (\ref{13})  appears as a necessary condition,
but it is not sufficient. Additionally, 
the compactification radius should be greater than its critical value.
Thus, we consider this quantization condition as a conjeture and 
we have applied the Lanczos method to test it.


\section{Numerical Results: The Lanczos Method} 

Now we present our  numerical
studies of 2-leg ladders with a number of sites
$N=2\times L$ where $L=4,6,8,10$ (due to the constraint of
periodic BC's). We apply Lanczos to each $S^z$-sector of the Hilbert
space from $S^z=0$ to $S^z={N\over2}(S_1+S_2)$.
We always set $J=1$ and make $J'$ range both in
AF and Ferromagnetic regimes. The dimerization parameter $\gamma$
is also varied for both CBA and SBA patterns.

\noindent 
Firstly, we consider AF couplings ($J'>0$) and 
vanishing dimerization ($\gamma=0$). In Fig.\ref{fig1} we show 3 plots
of the magnetization curve for weak $J'=0.2$, intermediate $J'=1.0$ and
strong coupling $J'=2.0$. For weak coupling we find two plateaux at 
$m=1/3,1$ and values of $h_{c1},h_{c2}=h_{sat}$ very close to the pure 1d
ferrimagnet. This amounts to a  smooth evolution from the one-chain case.
Moreover, at strong and intermediate couplings, the same picture holds true
which supports our effective Hamiltonians (\ref{9}),(\ref{10}) based on the
low-energy analysis. We have also checked numerically that the value of the saturation
field $h_{sat}$ fits nicely  the SWT result in eq. (\ref{7}), as clearly
shown in Fig.\ref{fig1}.
Another remarkable aspect of our plots is that they show a square-root
behaviour of the magnetization $m-m_c \sim \sqrt{h^2-h^2_c}$ 
in the transient region near the critical points,
as is also the case for standard AF spin-1/2 ladders \cite{c-h-p}.
Next, as a check of our numerical method, we consider SBA dimerization
with $\gamma=1$ and $J'=2.0$. In this situation the 2-leg ladder degenerates
onto a one-chain ferrimagnet with coupling $J=2.0$ and in Fig.\ref{fig2}
we reproduce exactly the existence of two plateaux and the value of the
saturation field $h_{sat}=3$ (in units $J=2$.) We find that this SBA
picture is robust by plotting the case $\gamma=0.5$.

\noindent In order to see the effect of different dimerization patterns
in the formation of the magnetization plateaux, we have also studied the
CBA dimerization configuration. We plot the case CBA for $J'=2.0$ 
and $\gamma=0.5, 0.8$ in Fig.\ref{fig3}.
Interestingly enough, {\em we clearly find a
new plateau at $m=2/3$} 
in addition to the previous ones at $m=1/3$ and $m=1$.
This agrees with eq. (\ref{13}) with $T_u=2$.
The width of this new plateau grows with increasing dimerization.
The new plateau shows the appearance of an energy gap between the two 
energy bands in the spectrum of CBA configuration.
This picture is true for CBA configuration ($J'>0$) when $\gamma \neq 1$.
By setting $\gamma=1$ the ladder will be decomposed to isolated
4-sites plaquette which does not reflect the properties of a system
in the thermodynamic limit ($N \rightarrow \infty$).

\noindent 
Secondly, we consider Ferromagnetic  couplings ($J'<0$) with
different dimerizations. In this case the CBA pattern is more 
interesting for we have found that it exhibits two phases, one gapless
for $\gamma<\gamma_c$ and another gapped for $\gamma>\gamma_c$, 
in the $J'-\gamma$ diagram with a phase-separating 
line which goes as  $\gamma_c \sim 0.42$ 
when $J'\rightarrow -\infty$ \cite{QRG}.

\begin{figure}
\hspace{-0.8cm}
\epsfxsize=8.5cm \epsfysize=8cm  \epsffile{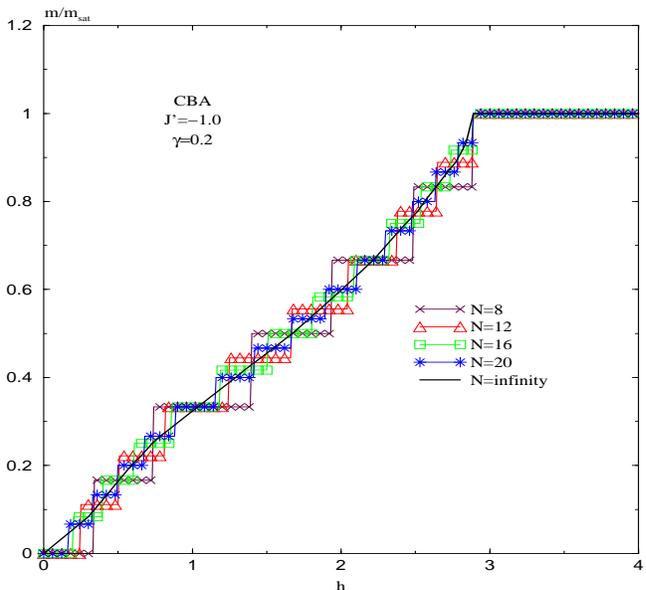}
\narrowtext
\caption[]{Magnetization curves for ferro CBA ladders with  $J'=-1.0$,
$\gamma=0.2$. There is no plateau which reveals no gap in the full spectrum.
}
\label{fig4} 
\end{figure}

In Fig.\ref{fig4} we consider a plot for $J'=-1.0$ and $\gamma=0.2$
which  happens to be in the gapless phase. 
Accordingly with this behaviour, there is no plateau at $m=0$
and the transient region immediately starts at $h=0$.
Moreover, the previously robust $1/3$-magnetization plateau is now
absent. The reason for this is because now the GS has no longer total
spin ${N\over2}(S_1-S2)$ but $0$.
This is completely in agreement with the effective Hamiltonian which has 
been obtained for this region in \cite{QRG}. The effective Hamiltonian for
CBA configuration ($J'<0$) and $\gamma < \gamma_c$ is a bond-alternation
spin 3/2 AFH model in the gapless phase \cite{QRG}. It has been shown that
the gapless phase of spin 3/2 AFH chain is in the universality class of a 
spin 1/2 AFH chain\cite{yamamoto}. As it has been noted in the introduction
the model has {\it no} plateau in the magnetization curve up to the 
saturation field.

\begin{figure}
\hspace{-0.8cm}
\epsfxsize=8.5cm \epsfysize=8cm  \epsffile{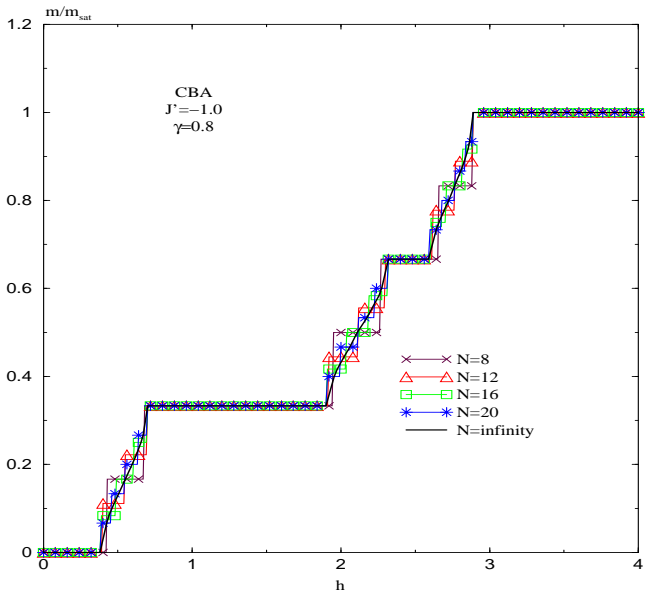}
\narrowtext
\caption[]{Magnetization plateaux for ferro CBA configurations with $J'=-1.0$,
$\gamma=0.8$. Four plateaux appear in this region at $m=0\;,1/3\;,2/3\;,1$.
}
\label{fig5} 
\end{figure}

Finally, we consider the opposite case of parameters corresponding to
the gapped phase, namely $J'=-1.0$ and $\gamma=0.8$, in Fig.\ref{fig5}.

This plot shows a very rich structure. It shows four magnetization plateaux
at $m=0,1/3,2/3,1$.  Since the model has an energy gap 
in this region the $m=0$ plateau extends until the magnetic field 
closes the gap in the spectrum 
 $h_c=\Delta$. More interesting is now the $m=1/3$ plateau,
{\em which is not natural for now the GS is not ferrimagnetic when $J'<0$.}
In fact, this plateau does not start at $h=0$ 
as in Figs.\ref{fig1}-\ref{fig3}. 
To explain the nature of these
plateaux we can resort to a strong coupling RG analysis and map
the original Hamiltonian (\ref{1}) onto a AFH one-chain of spins $3/2$
with dimerization (see eq.(5) in \cite{QRG}) 
and apply the OYA argument.
We have also checked that the saturation field $h_{sat}$ is also correctly
predicted by SWT in eq. (\ref{14}).

The situation of SBA at $J'<0$ is similar to Fig.\ref{fig4}. This can 
be understood by the effective low-energy Hamiltonian which is
a homogenous spin 3/2 AFH chain\cite{QRG}. The model is gapless for 
the whole range of magnetic field which results to no plateaux.


\noindent


\section{Conclusions}

We have addressed the study of alternating $(S_1,S_2)$-spin ladders
in the presence of an external uniform magnetic field in a fixed direction.
The low energy properties of 
this problem is relevant in  understanding the behaviour of ferrimagnetic 
systems in magnetic fields as compared with the properties exhibited by
more standard antiferromagnetic ladders.

In summary, we have shown that mixed-spin systems
show extra magnetization plateaux in addition to the
natural ones at $m=1/3 \ \mbox{and}\ 1$ and we have
presented a complete account for the diversity of 
plateaux-patterns when we vary the interchain  
and  dimerization couplings.

We have undertaken a complete numerical study using the Lanczos method
of the 2-leg ladders with alternating spins $S_1=1/2, S_2=1$, for 
several regimes of the interchain coupling constant $J'$, both antiferro-
and ferromagnetic, and  for several types of dimerization patterns 
(CBA and SBA).
In the antiferromagnetic regime ($J'>0$) the SBA configuration shows only
the natural plateaux at  $m=1/3 \ \mbox{and}\ 1$ while the CBA configuration
shows an extra plateau at $m=2/3$. In the ferromagnetic regime ($J'<0$)
the SBA configuration shows no plateau where the model remains gapless from
the zero field up to the saturation field. The CBA configuration 
in the $J'<0$ regime has more interesting features. For $\gamma < \gamma_c$
the model is gapless for all range of magnetic field and results to no plateau,
while for $\gamma > \gamma_c$ the model is gapful. Then the first plateau
appears at $m=0$. There are also two other plateaux at $m=1/3 \ \mbox{and}\ 2/3$
which are expressed by the dimerized $S$=3/2 AFH effective model. The trivial
plateau at $m=1$ counts the number of plateaux to 4 in this case.

The magnetization curves obtained in this way provide a whole picture of 
the response of mixed-spin systems to external magnetic fields.

We have complemented these  numerical studies with additional analytical
methods such as Spin Wave Theory (in two fashions), effective low-energy
mappings of the type used in Quantum RG methods and Abelian Bosonization.
Although none of these analytical methods provide a complete understanding
of the problem, they however give us some useful partial knowledge 
that fits nicely into our findings obtained with numerical methods.

Concerning the possible experimental realization of the magnetic ferriladders
studied in this work, we would like to point out that the field of 
organic chemistry is a likely candidate for synthesizing this type of
materials \cite{canishi,hosokoshi,nishizawa}.

{\bf Acknowledgements} 
We acknowledge discussions with  
M. Abolfath,  H. Hamidian and G. Sierra,
and the Centro de Supercomputaci\'on Complutense for the allocation
of CPU time in the SG-Origin 2000 Parallel Computer
and Max-Planck-Institut f\"ur Physik komplexer Systeme
for computer time.
M.A.M.-D. was supported by the DGES spanish grant
PB97-1190. 

\vspace{-0.5 true cm}


\begin{thebibliography}{99}
\bibitem{comment1}
We have performed the same QRG analysis for a one-chain finding the
same picture of two plateaux at $m=1/3,1$ but with different values
of $h_{c1},h_{c2}$. This agrees with numerical results.
\end{thebibliography}
\end{document}